\documentclass[aps,superscriptaddress,twocolumn,a4paper,showpacs]{revtex4}
\usepackage{graphicx}
\usepackage[all]{xy}
\usepackage{amsmath}
\usepackage{amssymb}
\usepackage{tensor}
\newcommand{\be}{\begin{equation}}
\newcommand{\ee}{\end{equation}}
\newcommand{\ben}{\begin{eqnarray}}
\newcommand{\een}{\end{eqnarray}}
\newcommand{\bes}{\begin{subequations}}
\newcommand{\ees}{\end{subequations}}
\def\bal#1\eal{\begin{align}#1\end{align}}

\newcommand{\bb}{\bibitem}

\newcommand{\LL}{{\cal L}}

\newcommand{\vphi}{{\varphi}}

\begin{document}

\title{Compact Chern-Simons vortices}
\author{D. Bazeia}\email{bazeia@fisica.ufpb.br}\affiliation{Departamento de F\'\i sica, Universidade Federal da Para\'\i ba, 58051-970 Jo\~ao Pessoa, PB, Brazil}
\author{L. Losano}\email{losano@fisica.ufpb.br}\affiliation{Departamento de F\'\i sica, Universidade Federal da Para\'\i ba, 58051-970 Jo\~ao Pessoa, PB, Brazil}
\author{M.A. Marques}\email{mam@fisica.ufpb.br}\affiliation{Departamento de F\'\i sica, Universidade Federal da Para\'\i ba, 58051-970 Jo\~ao Pessoa, PB, Brazil}
\author{R. Menezes}\email{rmenezes@dce.ufpb.br}
\affiliation{Departamento de Ci\^encias Exatas, Universidade Federal
da Para\'{\i}ba, 58297-000 Rio Tinto, PB, Brazil}
\affiliation{Departamento de F\'\i sica, Universidade Federal de Campina Grande, 58109-970, Campina Grande, PB, Brazil}
\date{\today}
\begin{abstract}
We introduce and investigate new models of the Chern-Simons type in the three-dimensional spacetime, focusing on the existence of compact vortices. The models are controlled by potentials driven by a single real parameter that can be used to change the profile of the vortex solutions as they approach their boundary values. One of the models unveils an interesting new behavior, the tendency to make the vortex compact, as the parameter increases to larger and larger values. We also investigate the behavior of the energy density and calculate the total energy numerically.
\end{abstract}

\maketitle
\section{Introduction}
Vortices are topological structures that appear in $(2,1)$ spacetime dimensions, under the action of a complex scalar field coupled to a gauge field which evolves obeying the local $U(1)$ symmetry. The study of relativistic vortices started in Ref.~\cite{novortex,vega}, by considering a model governed by the Maxwell dynamics. However, one can also investigate vortices with the dynamics controlled by the Chern-Simons term, whose general properties where shown in Refs.~\cite{cs,jac,csp1}. The first studies considering Chern-Simons vortices appeared in Refs.~\cite{cs1,cs2,cs3}; for more on this see, e.g., Ref.~\cite{dunne}. 

The class of vortices that appears in models driven by the Chern-Simons term presents some interesting features, beyond the fact that they engender a magnetic flux that is quantized by its intrinsic vorticity. A peculiar behavior which deserves to be pointed out is the existence of an electric field and the fact that the magnetic field is related to the electric charge density associated to the Chern-Simons vortex solutions. 

Soon after the first works on Chern-Simons vortex \cite{cs1,cs2,cs3}, other generalized models that also support vortex solutions appeared in Refs.~\cite{gen1,gen2}. On the other hand, over the past twenty years generalized models have been considered with other motivations. One of the first models with non-canonical kinetic term was presented in Ref.~\cite{kinf}, in the context of inflation. Afterwards, Refs.~\cite{cosm1,cosm2} studied generalized models as a tentative to solve the cosmic coincidence problem \cite{cosm1,cosm2}. It is worth mentioning that non-canonical models may present distinct features from the standard case. In the inflation scenario, for instance, these models may not require the presence of a potential to drive inflation. One can also study generalized models with defect structures in field theories, as in Refs.~\cite{babichev1,babichev2}. These models have been studied in several papers, in particular in \cite{kd1,kd2,kd3,kd4,kd5,kd6,kd7,cas}. An interesting fact here is that compact structures, firstly considered in Ref.~\cite{rosenau} in models comprising nonlinearity and nonlinear dispersion, and later studied in Refs.~\cite{ck1,ck2,ck3,ck4}, are also found in generalized models, as seen in Ref.~\cite{ckg}. Compact vortices in generalized Maxwell-Higgs models were found in Ref.~\cite{cvmh}. They engender interesting features, such as the mapping of an infinitely long solenoid, if one includes a third spatial dimension.

In this work, we want to further investigate the presence of vortices in Chern-Simons models, but now focusing on generalized models. The idea is to follow Ref.~\cite{bazeiacs} and propose new solutions of the vortex-like type in the case of non-canonical models. To do this, in Sec.~\ref{model} we first review the properties of the generalized model presented in Ref.~\cite{bazeiacs}, and then go on and investigate two distinct models in Sec.~\ref{exr}. The first model presents a parameter that controls the potential, which may be increased to make the model become the standard model described in Refs.~\cite{cs1,cs2}. The second model is also controlled by a real parameter, but now it starts at unit value with the standard model of Refs.~\cite{cs1,cs2} and makes the vortex compact if it is increased to larger and larger values. We conclude the work in Sec.~\ref{conclusions}.
 
%%%%%%%%%%%%%%%%%%%%%%%%%%%%%%%
\section{Generalized Chern-Simons Vortices}\label{model}
We work with a complex scalar field and a gauge field governed by the pure Chern-Simons dynamics in $(2,1)$ spacetime dimensions. We use the class of generalized models presented in Ref.~\cite{bazeiacs}, which is described by the action $S=\int d^3x\LL$, with the Lagrangian density $\LL$ given by
\begin{equation}\label{lcomp}
\LL = \frac{\kappa}{4}\epsilon^{\alpha\beta\gamma}A_\alpha F_{\beta\gamma} + K(|\vphi|)\overline{D_\mu \vphi}D^\mu\vphi -V(|\vphi|).
\end{equation}
In the above expression, $\vphi$ is the complex scalar field,  $D_{\mu}=\partial _{\mu }+ieA_{\mu }$, $F_{\mu\nu}=\partial_\mu A_\nu-\partial_\nu A_\mu$ and $V(|\vphi|)$ is the potential. Also, $\kappa$ has to be a constant, to keep gauge invariance of the action. The dimensionless function $K(|\vphi|)$ is in principle arbritrary, although it has to allow the existence of solutions with finite energy. The standard case studied in Ref.~\cite{cs2} is obtained for $K(|\vphi|)=1$. We are using $A^\mu = (A^0,\vec{A})$, such that the electric and magnetic fields are defined by
\be\label{eb}
E^i = F^{i0} = -\dot{A}^i - \partial_i A^0 \quad\text{and}\quad B = -F^{12},
\ee
with $(E_x,E_y)\equiv E^i$ and $i=1,2$. The equations of motion for the fields $\vphi$ and $A_\mu$ are given by
\bes\label{gcseom}
\begin{align}
 D_\mu (K D^\mu\vphi)&= \frac{\vphi}{2|\vphi|}\left(K_{|\vphi|}\overline{D_\mu \vphi}D^\mu\vphi -V_{|\vphi|} \right), \\ \label{cseqs}
 \frac{\kappa}{2} \epsilon^{\lambda\mu\nu}F_{\mu\nu} &= J^\lambda,
\end{align}
\ees
where the current is $J_\mu = ieK(|\vphi|)(\bar{\vphi}D_\mu \vphi-\vphi\overline{D_\mu\vphi})$. The energy momentum tensor $T_{\mu\nu}$ for the generalized model \eqref{lcomp} is given by 
\bal
T_{\mu\nu}&=K(|\vphi|)\left( \overline{D_\mu \vphi}D_\nu \vphi + \overline{D_\nu \vphi}D_\mu \vphi\right) \nonumber\\
          &\hspace{4mm}- \eta_{\mu\nu} \left( K(|\vphi|)\overline{D_\lambda \vphi}D^\lambda\vphi -V(|\vphi|) \right).
\eal

We now consider static configurations. In this case, the energy density is given by
\be\label{gcsrho}
\rho\equiv T_{00} = 2 e^2 K(|\vphi|) A_0^2 |\vphi|^2 \!+\! V(|\vphi|) \!-\! K(|\vphi|)\overline{D_\lambda \vphi}D^\lambda\vphi,
\ee
The component $A_0$ that appears in the above equations is not an independent function.  From the temporal component of the Eq.~\eqref{cseqs}, we get that $A_0$ is constrained to obey
\be\label{A0}
A_0 = \frac{\kappa}{2e^2} \frac{B}{|\vphi|^2K(|\vphi|)}.
\ee
This leads to
\be\label{rhob}
\rho\equiv T_{00} = \frac{\kappa^2}{4e^2} \frac{B^2}{|\vphi|^2K(|\vphi|)} + K(|\vphi|)\overline{D_i \vphi}D_i\vphi + V(|\vphi|).
\ee
To search for vortexlike solutions, we take the usual ansatz
\bes\label{ansatz}
\begin{align}
\vphi(r,\theta) &=g(r)e^{i n\theta},\\
A_0 &=A_0(r),\\
A_i &=-\epsilon_{ij} \frac{x^j}{er^2}[a(r)-n]
\end{align}
\ees
where $r$ and $\theta$ are polar coordinates and $n$ is an integer, the vortex winding number. The functions $g(r)$ and $a(r)$ must obey the boundary conditions
\be\label{bcond}
g(0) = 0, \quad a(0)= n, \quad \lim_{r\to\infty} g(r) = v, \quad \lim_{r\to\infty} a(r) = 0.
\ee
With the ansatz \eqref{ansatz}, we have $\overline{D_\mu \vphi}D^\mu\vphi=e^2g^2A_0^2-({g^\prime}^2+a^2g^2/r^2)$, where the prime denotes the derivative with respect to $r$. Furthermore, the electric and magnetic fields in Eq.~\eqref{eb} become
\be\label{ebcs}
E^i=-A_0^\prime \frac{x^i}{r} \quad\text{and}\quad B = -\frac{a^\prime}{er}.
\ee
The magnetic flux $\Phi=2\pi\int_0^\infty r dr B(r)$ is quantized:
\be\label{mflux}
\Phi=\frac{2\pi n}{e}.
\ee
The electric charge is given by $Q =2\pi\int rdr J^0$. By using Eq.~\eqref{A0}, one can show that the charge can be written in terms of the magnetic flux \eqref{mflux} as $Q = -\kappa \Phi$, which makes the electric charge to be quantized. The equations of motion \eqref{gcseom} with the ansatz \eqref{ansatz} are given by
\bes\label{eomcsansatz}
\begin{align}
\frac{1}{r} \left(rK g^\prime\right)^\prime + K g \left(e^2 A_0^2-\frac{a^2}{r^2} \right)+& \nonumber\\
+ \frac12 \left(\left(e^2g^2A_0^2-{g^\prime}^2-\frac{a^2g^2}{r^2}\right)K_{|\vphi|} -V_{|\vphi|} \right) &= 0, \\ \label{a0csansatz}
 \frac{a^\prime}{r} + \frac{2K e^3g^2 A_0}{\kappa} &= 0, \\
 {A_0^\prime} + \frac{2K ea g^2}{\kappa r} &= 0.
\end{align}
\ees
Also, the energy density \eqref{rhob} becomes
\be\label{rhoans}
\rho =\frac{\kappa^2}{4e^4} \frac{{a^\prime}^2}{r^2 g^2K(g)} + \left({g^\prime}^2+\frac{a^2g^2}{r^2}\right)K(g) + V(g).
\ee

To find the energy density \eqref{rhoans}, one has to solve the equations of motion \eqref{eomcsansatz}, which is a very complicated task, since they are coupled second order differential equations. However, in Ref.~\cite{bazeiacs}, it was shown that the first order equations
\be\label{focs}
g^\prime = \frac{ag}{r} \quad\text{and}\quad a^\prime =-\frac{2e^2rg}{\kappa}\sqrt{KV}  
\ee
solve the equations of motion \eqref{eomcsansatz} if the functions $K(|\vphi|)$ and $V(|\vphi|)$ are constrained to obey
\be\label{constraintVK}
\frac{d}{dg} \left(\sqrt{\frac{V}{g^2K}}\,\right) = -\frac{2e^2}{\kappa}gK.
\ee

Below we further explore generalized models of the class presented above, searching for vortex solutions that solve the first order equations \eqref{focs} and obey the constraint \eqref{constraintVK}. For simplicity, from now on we consider $e=\kappa=v=1$ and work with dimensionless fields, setting $n=1$ to explore the case of unit vorticity.

%%%%%%%%%%%%%%%%%%%%%%%%%%%%%
\section{New models}\label{exr}

We see from \eqref{lcomp} that the model is specified once the two functions $K(|\varphi|)$ and $V(|\varphi|)$ are given explicitly. Below we study two distinct possibilities.

\subsection{Model 1}

\begin{figure}[t!]
\centering
\includegraphics[width=6cm]{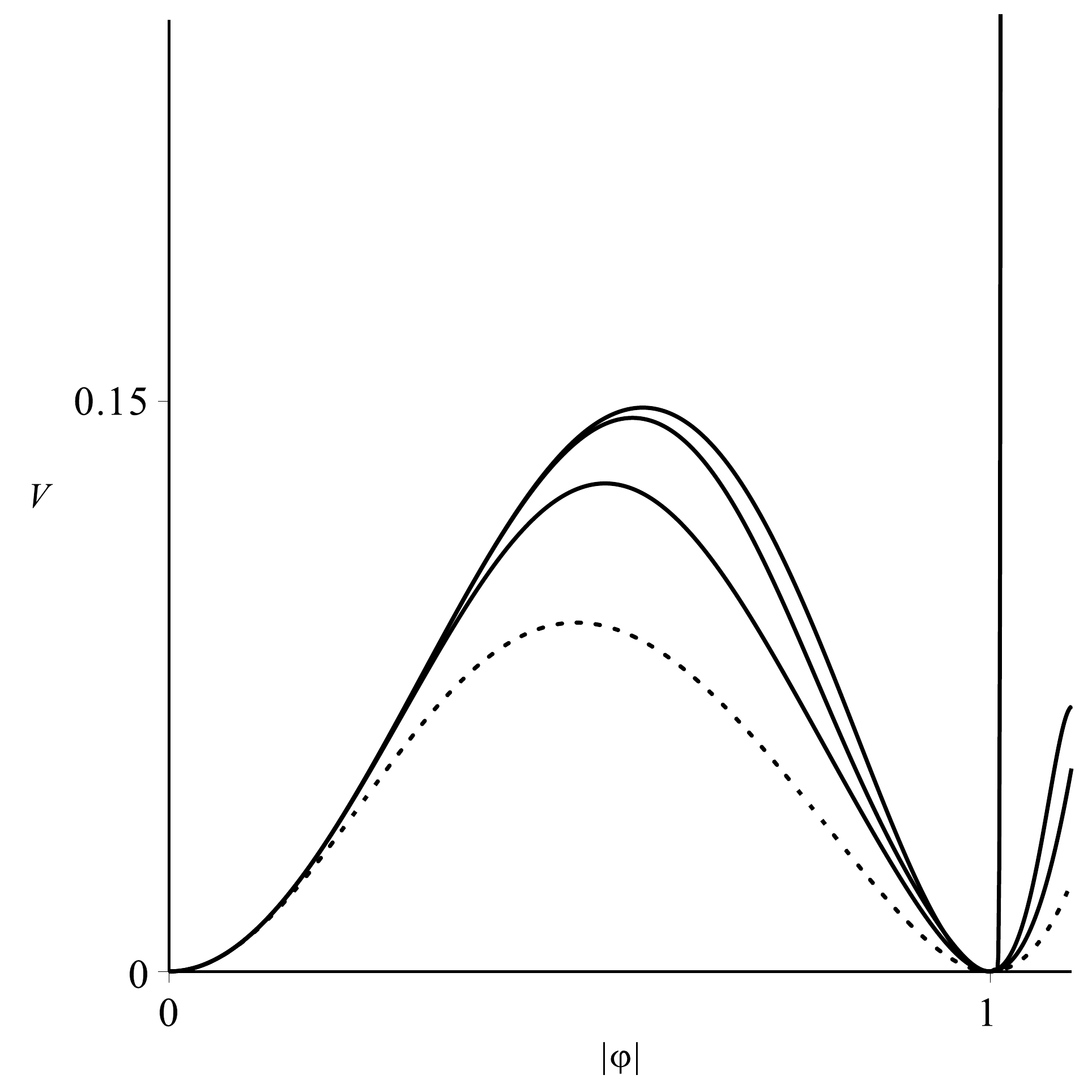}
\caption{The potential of Eq.~\eqref{potrob} is depicted for $l=1,2,4$ and $64$, with the case $l=1$ as the dotted line.}
\label{figVr}
\end{figure}

Let us first consider the model that is specified by the functions
\bes\bal
K(|\vphi|) &= \left(1-|\vphi|^{2l}\right)^2, \\
V(|\vphi|) &= |\vphi|^2\left(1-|\vphi|^{2l}\right)^2 \nonumber\\ \label{potrob}
           &\hspace{4mm}\times\left(1-|\vphi|^2 + \frac{2}{l+1}|\vphi|^{2l+2} -\frac{1}{2l+1}|\vphi|^{4l+2} \right)^2,
\eal\ees
where $l$ is a real parameter such that $l\geq1$. The above pair of functions is compatible with Eq.~\eqref{constraintVK}. This potential has a minimum at $|\vphi|=0$ and $|\vphi|=1$, regardless the value of $l$. Whilst these minima are fixed, the maximum between them, which cannot be calculated analytically, changes with $l$. The above potential also has other minimum for $|\vphi|>1$, which is solution of the equation
\be
1-|\vphi|^2 + \frac{2}{l+1}|\vphi|^{2l+2} -\frac{1}{2l+1}|\vphi|^{4l+2} =0.
\ee
However, here we are only interested to find solutions in the sector $0\leq |\vphi|\leq 1$. To illustrate this, in Fig.~\ref{figVr} we depict the potential \eqref{potrob} in the interval $0\leq|\vphi|\leq1+\delta$, for $\delta$ small and for several values of $l$, to illustrate how the potential behaves around the minimum at $|\varphi|=1$. One can see that the concavity of the potential in the minimum $|\vphi|=1$ narrows as we increase $l$. However, for $\vphi\to1$ from the left, the potential has almost the same behavior. Also, the minimum located outside of the aforementioned sector gets closer to $|\vphi|=1$ as $l$ gets larger. In the limit $l\to\infty$, this potential tends to become
\be\label{vinfr}
V_\infty(r)=
\begin{cases}
|\vphi|^2\left(1-|\vphi|^{2}\right)^2,\,\,\,& |\vphi|<1\\
\infty, \,\,\, & |\vphi|>1.
\end{cases}
\ee
Therefore, inside the sector $0\leq|\vphi|\leq1$, the potential tends to behave as the one studied in Refs.~\cite{cs1,cs2}.
%%%%%%%%%%%%%%%%%%%%%%%
%%%%%%%%%%%%%%%%%%%%%%

\begin{figure}[t!]
\centering
\includegraphics[width=6cm]{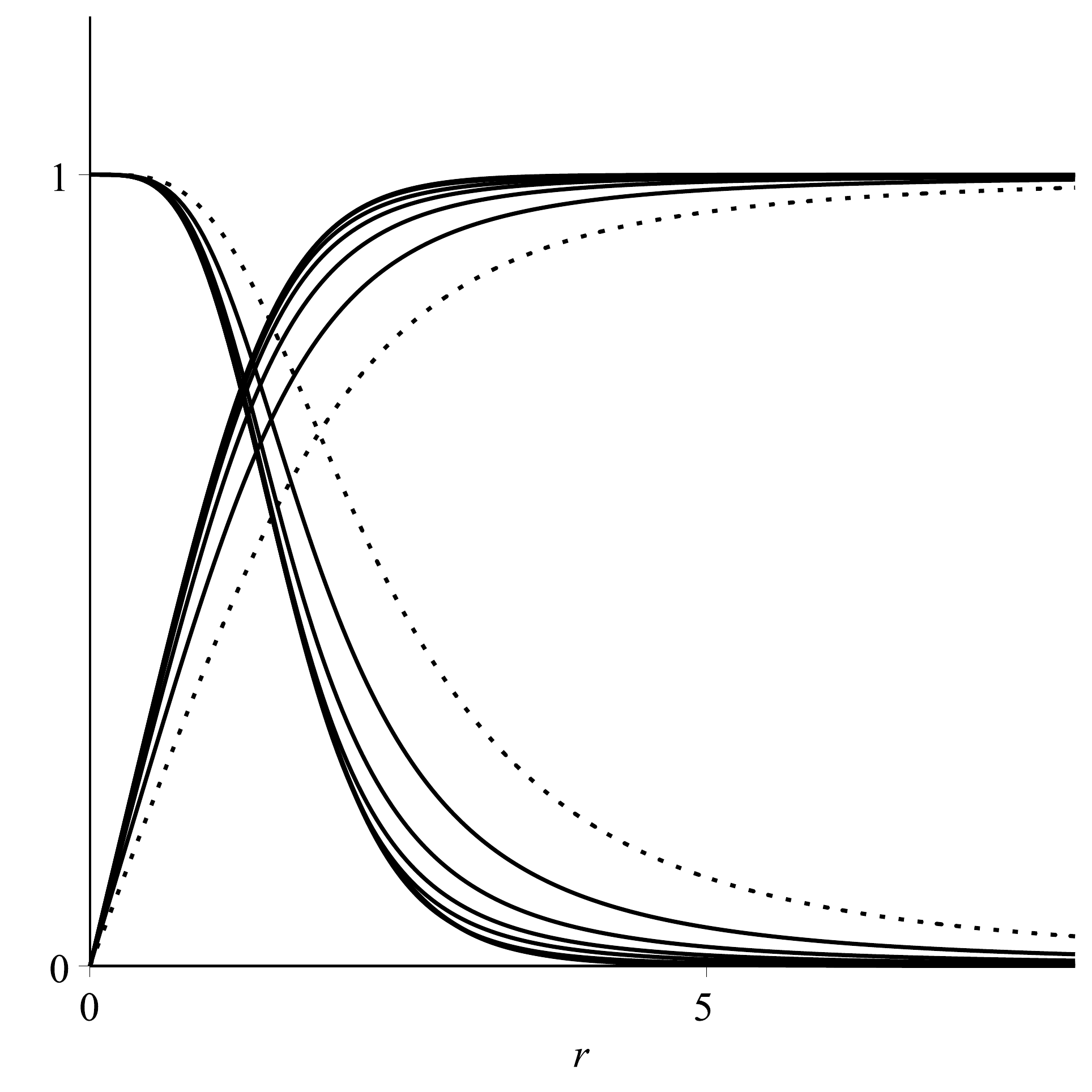}
\caption{The solutions $a(r)$ (descending lines) and $g(r)$ (ascending lines) of Eqs.~\eqref{foroba} and \eqref{forobb}, depicted for $l=1, 2, 4, 8, 16, 64, 256$ and $1024$, with the case $l=1$ as the dotted line.}
\label{figsolr}
\end{figure}

In the general case, the first order equations \eqref{focs} become
\bes\bal
g^\prime &= \frac{ag}{r}\label{foroba} \\
a^\prime &=-2r g^2\left(1-{g}^{2l} \right)^2\left(1 - g^2 + \frac{2 g^{2l+2}}{l+1} - \frac{g^{4l+2}}{2l+1} \right)\label{forobb}.
\eal\ees
We can study the behavior of the above equations near the origin by considering $a(r) = 1-a_0(r)$ and $g(r) = g_0(r)$ up to first order in $a_0(r)$ and $g_0(r)$. This leads to $a_0(r)\propto r^{4}$ and $g_0(r) \propto r$. The first order equations \eqref{foroba} and \eqref{forobb} are very hard to be solved analytically, so we conduct the investigation numerically. In Fig.~\ref{figsolr}, we display the solutions of Eqs.~\eqref{foroba} and \eqref{forobb} for several values of $l$. Notice that, even though the solutions tend to approach their boundary condition as $l$ increases, they do not compactify, due to the fact that the potential admits the limit \eqref{vinfr}, which does not give rise to compact solutions inside the sector that is being considered.
%%%%%%%%%%%%%%%%%%%%%%%%%%
%%%%%%%%%%%%%%%%%%%%%%%%%%

%%%%%%%%%%%%%%%%%%%%%%%%%%%%%%%%%%%%
\begin{figure}[t!]
\centering
\includegraphics[width=4.25cm]{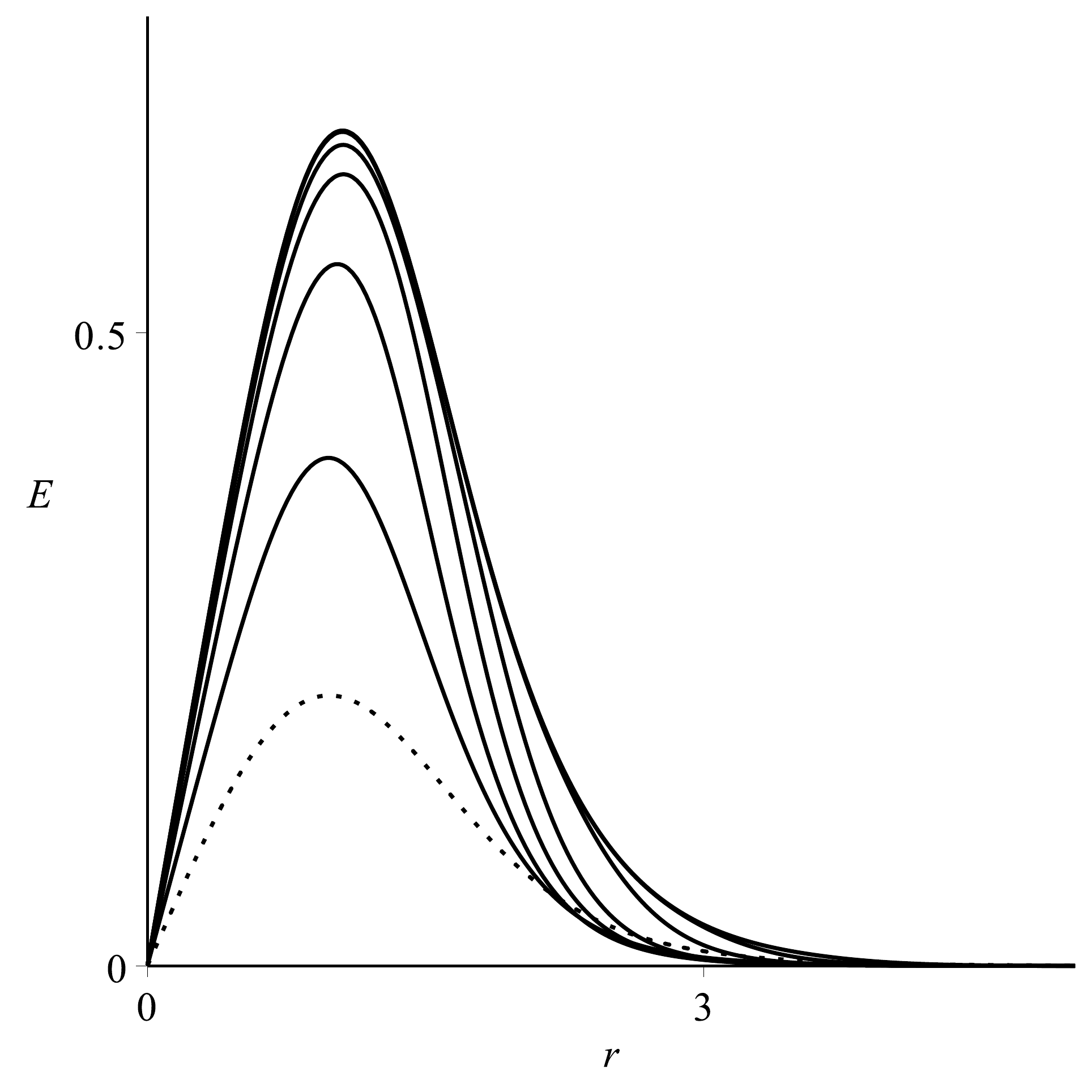}
\includegraphics[width=4.25cm]{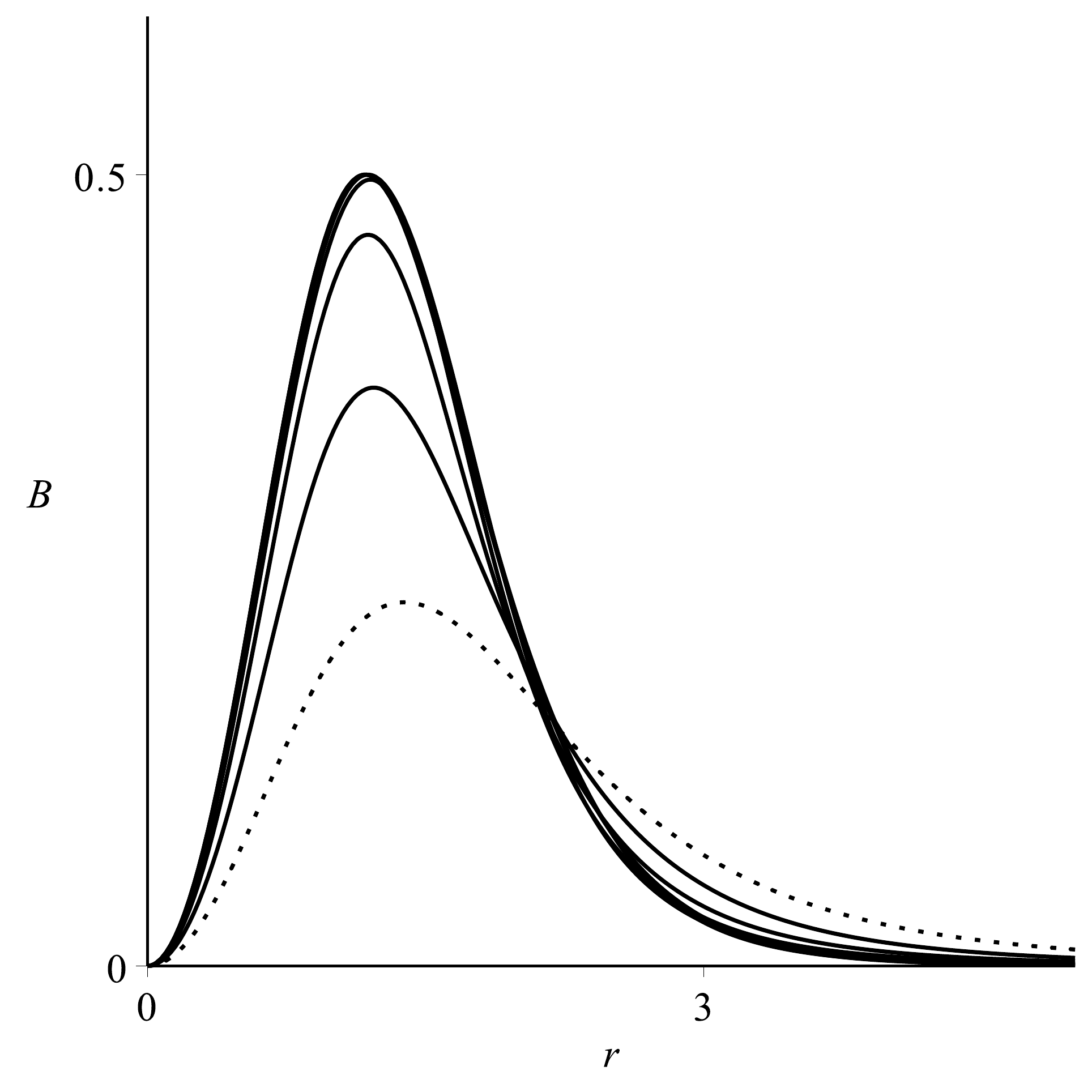}
\includegraphics[width=4.25cm]{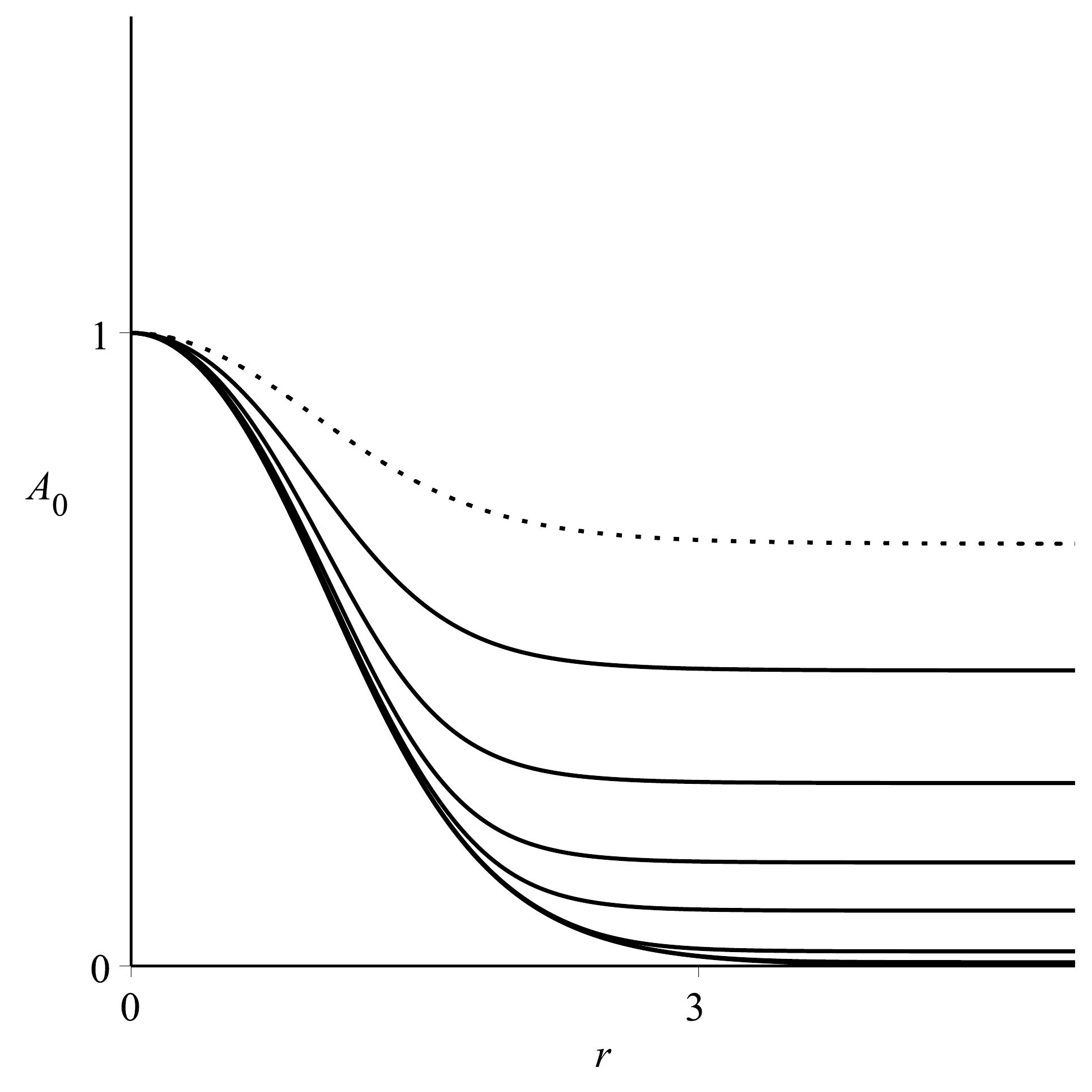}
\includegraphics[width=4.25cm]{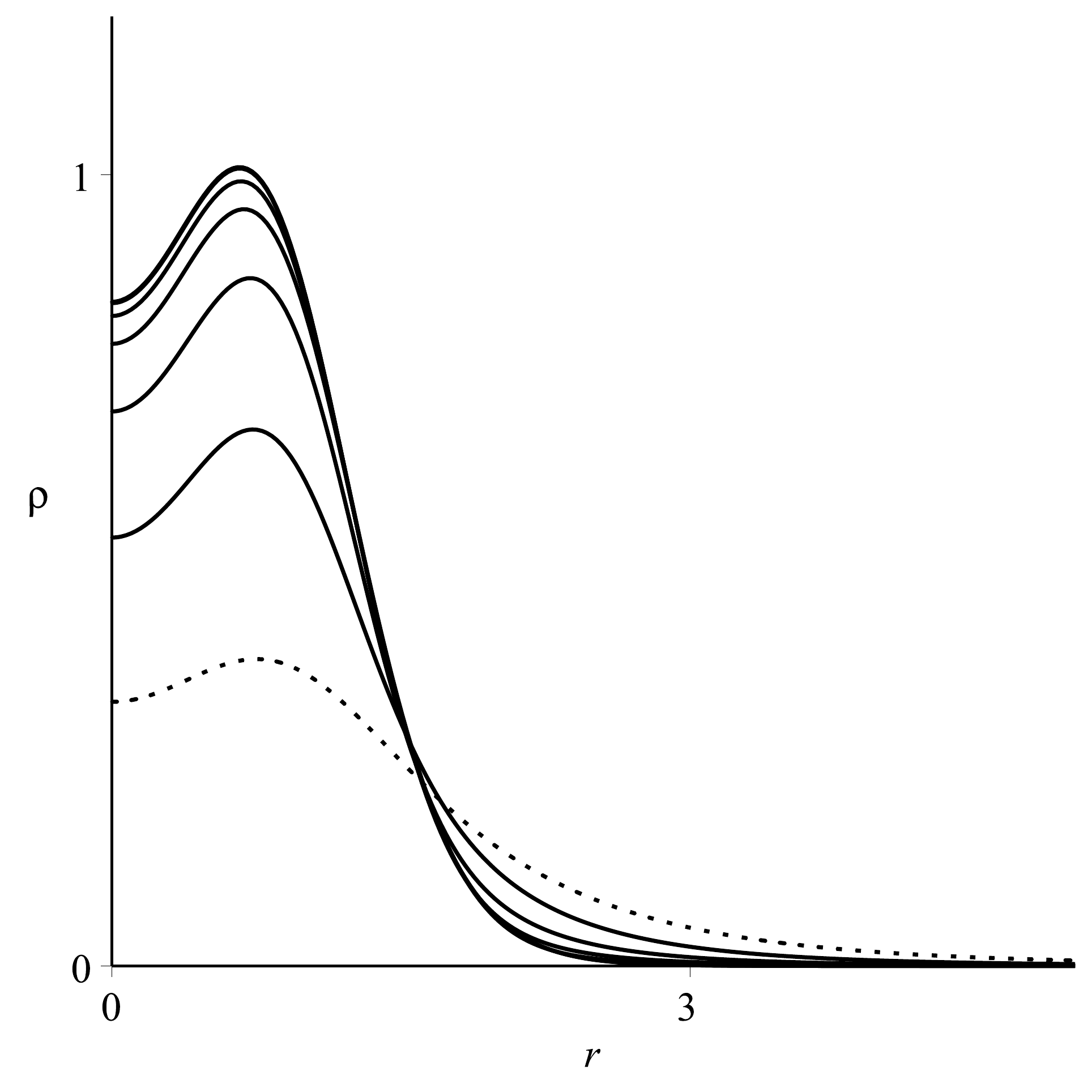}
\caption{The electric (top left) and magnetic (top right) fields \eqref{ebcs}, the temporal gauge field \eqref{A0} (bottom left) and the energy density \eqref{rhor} (bottom right) plotted for $l=1, 2, 4, 8, 16, 64, 256$ and $1024$, with the case $l=1$ as the dotted line.}
\label{figebr}
\end{figure}
%%%%%%%%%%%%%%%%%%%%%%%%%%%%%%%%%%%

The electric and magnetic fields are obtained from Eq.~\eqref{ebcs}, so we use the numerical solutions of Eqs.~\eqref{foroba} and \eqref{forobb} and plot them in Fig.~\ref{figebr} for several values of $l$. We also display the temporal gauge component \eqref{A0}. The energy density can be calculated from Eq.~\eqref{rhoans}. In this case, it is given by
\bal\label{rhor}
\rho &=\frac{{a^\prime}^2}{4r^2 g^2\left(1-{g}^{2l} \right)^2} + \left({g^\prime}^2+\frac{a^2g^2}{r^2}\right)\left(1-{g}^{2l} \right)^2 \nonumber\\
&\hspace{4mm} + g^2\left(1-g^{2l}\right)^2\left(1 - g^2 + \frac{2 g^{2l+2}}{l+1} - \frac{g^{4l+2}}{2l+1} \right)^2.
\eal
We have plotted the above energy density for the numerical solutions of Eqs.~\eqref{foroba} and \eqref{forobb} in Fig.~\ref{figebr} for several values of $l$. As $l$ gets larger and larger, the vortex energy density gets an internal structure. Numerical integration of these energy densities over all the space gives the energy $E\approx2\pi$, which is independent of $l$.

%%%%%%%%%%%%%%%%%%%%%%%
\begin{figure}[t!]
\centering
\includegraphics[width=6cm]{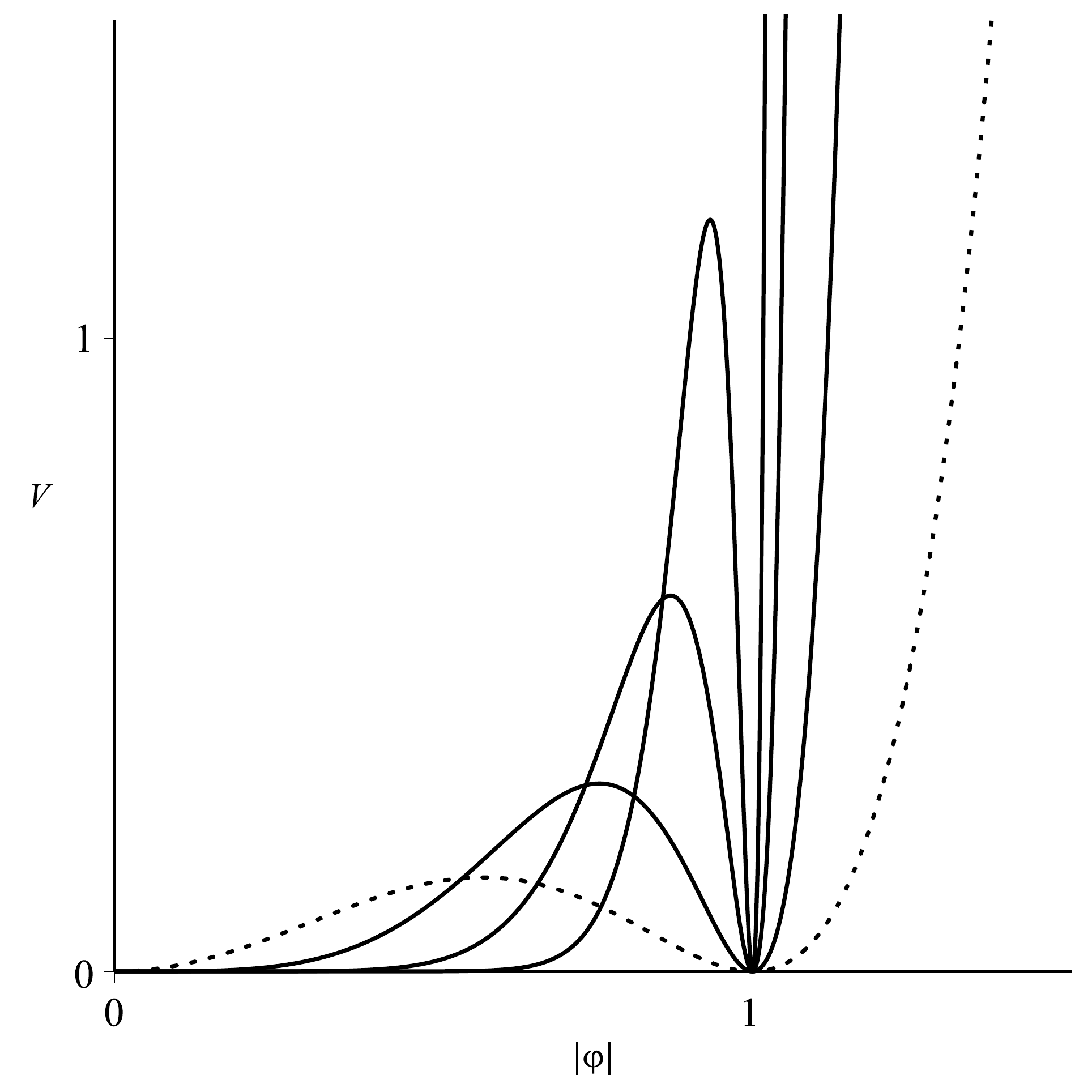}
\caption{The potential of Eq.~\eqref{potmat}, displayed for $l=1,2,4$ and $8$, with the case $l=1$ as the dotted line.}
\label{figVm}
\end{figure} 
%%%%%%%%%%%%%%%%%%%%%%

%%%%%%%%%%%%%%%%%%%%%%%%%%%%%%%%%%
\subsection{Model 2}\label{exm}

The second model which we consider is described by the generalized model \eqref{lcomp}, but is specified by the pair of functions
\be\label{potmat}
K(|\vphi|) = l{|\vphi|}^{2l-2} \quad\text{and}\quad V(|\vphi|) = l{|\vphi|}^{2l} \left(1 - {|\vphi|}^{2l}\right)^2,
\ee
which are compatible with the constraint \eqref{constraintVK}. In the above expressions, $l$ is a real parameter such that $l\geq1$. The model is simpler than the previous one. The potential has a minimum at $|\vphi|=0$ and $|\vphi|=1$. In between these two minima there is a local maximum at $|\vphi_{\max}|=1/3^{1/(2l)}$ such that $V(|\vphi_{\max}|) = 4l/27$. As $l$ increases, $|\vphi_{\max}|$ approaches more and more the minimum at $|\vphi|=1$, and $V(|\vphi_{\max}|)$ increases to lager and larger values. The standard case, which was studied in Refs.~\cite{cs1,cs2}, is obtained for $l=1$. In Fig.~\ref{figVm}, we display the potential \eqref{potmat} for several values of $l$ to illustrate its general behavior.

In this case, the first order equations \eqref{focs} become
\be\label{fomat}
g^\prime = \frac{ag}{r} \quad\text{and}\quad a^\prime =-2lr{g}^{2l}\left(1-{g}^{2l} \right).
\ee
Before trying to solve them, we can study their behavior near the origin by taking $a(r) = 1-a_0(r)$ and $g(r) = g_0(r)$ up to first order in $a_0(r)$ and $g_0(r)$. This shows that $a_0(r)\propto lr^{2(l+1)}/(l+1)$ and $g_0(r) \propto r$. A similar procedure can be used to find the asymptotic behavior of the functions by taking $a(r) = a_\infty(r)$ and $g(r) = 1-g_\infty(r)$ in Eqs.~\eqref{fomat}. This leads to $a_\infty(r) \propto 2lrK_1(2lr)$ and $g_\infty(r)\propto K_0(2lr)$. In the latter expressions, $K_m(z)$ is the modified Bessel function of the second kind. The asymptotic behavior shows that the functions $a(r)$ and $g(r)$ approach their respective boundary condition faster and faster as $l$ increases, a behavior that shows that the solutions tend to become compact for larger and larger values of $l$.

%%%%%%%%%%%%%%%%%%%%%%%
\begin{figure}[t!]
\centering
\includegraphics[width=6cm]{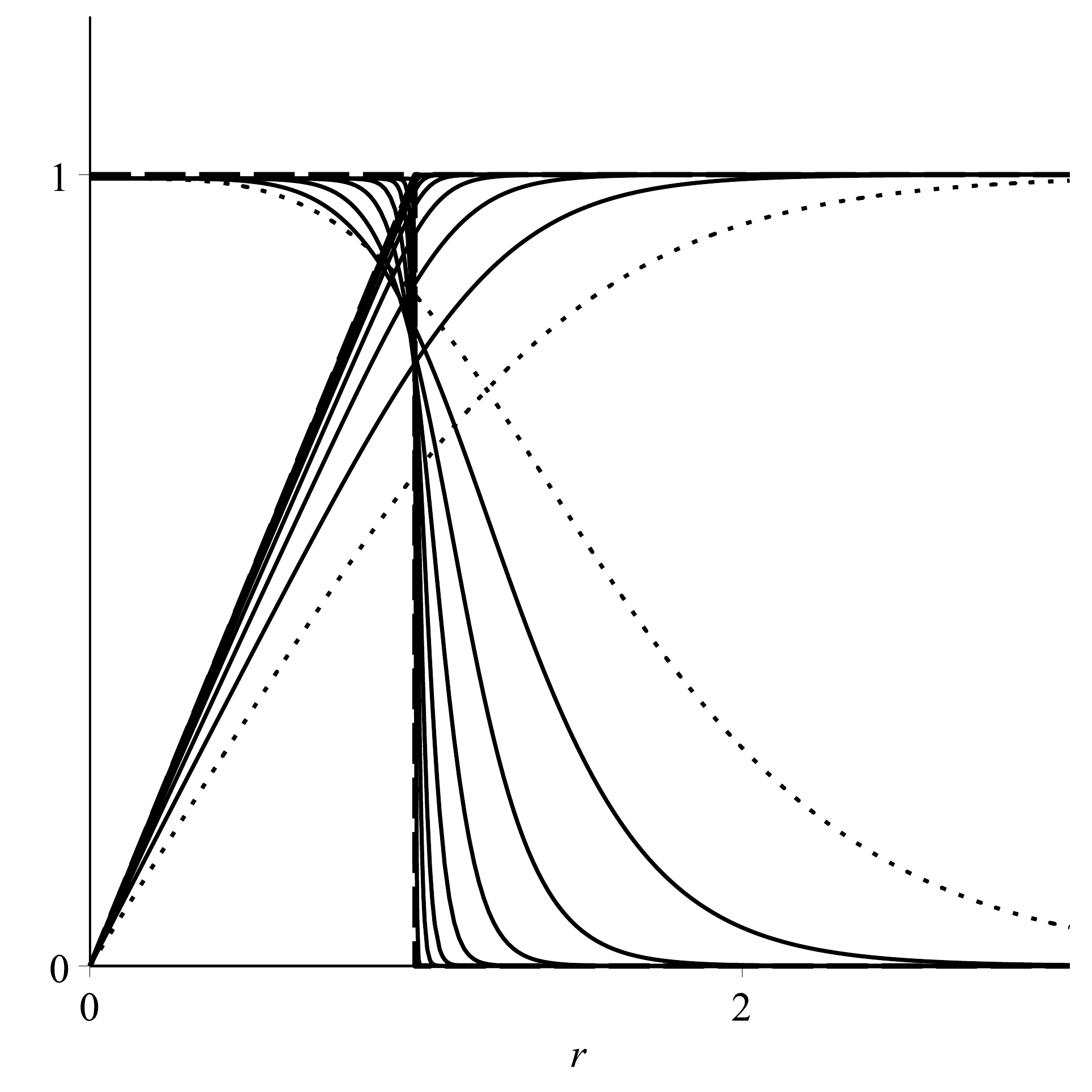}
\caption{The solutions $a(r)$ (descending lines) and $g(r)$ (ascending lines) of Eqs.~\eqref{fomat} plotted for $l=1$ and increasing to larger and larger values, with the case $l=1$ as the dotted line. The dashed lines stand for the compact limit \eqref{solcmat}.}
\label{figsolm}
\end{figure} 
%%%%%%%%%%%%%%%%%%%%%%

The first order equations \eqref{fomat} are very hard to be solved analytically for a general $l$. However, we found that, as $l$ increases, the solutions tend to get the compact profile
\bes\label{solcmat}
\ben
a_c(r)&=&
\begin{cases}
1,\,\,\,& r \leq r_c\\
0, \,\,\, & r>r_c,
\end{cases} \\
g_c(r)&=&
\begin{cases}
\frac{r}{r_c},\,\,\,& r \leq r_c\\
1, \,\,\, & r>r_c.
\end{cases}
\een
\ees
where $r_c$ is the compactification radius, which is calculated numerically for a very large value of $l$; it has the value $r_c\approx0.995$. In Fig.~\ref{figsolm}, we depict the solutions of Eqs.~\eqref{fomat} for several values of $l$, including the limit given by Eq.~\eqref{solcmat}.

%%%%%%%%%%%%%%%%%%%%%%%%
\begin{figure}[t!]
\centering
\includegraphics[width=4.26cm]{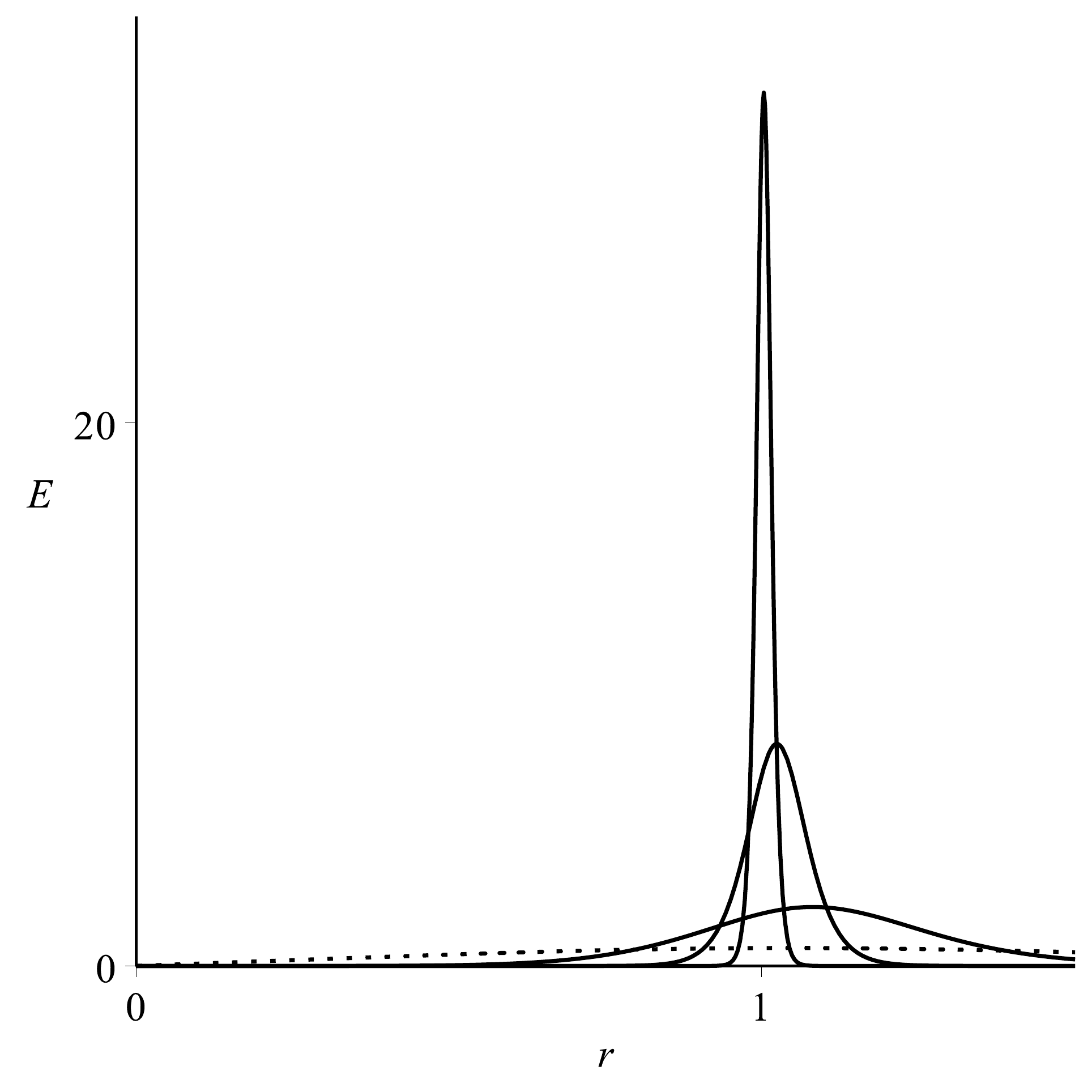}
\includegraphics[width=4.26cm]{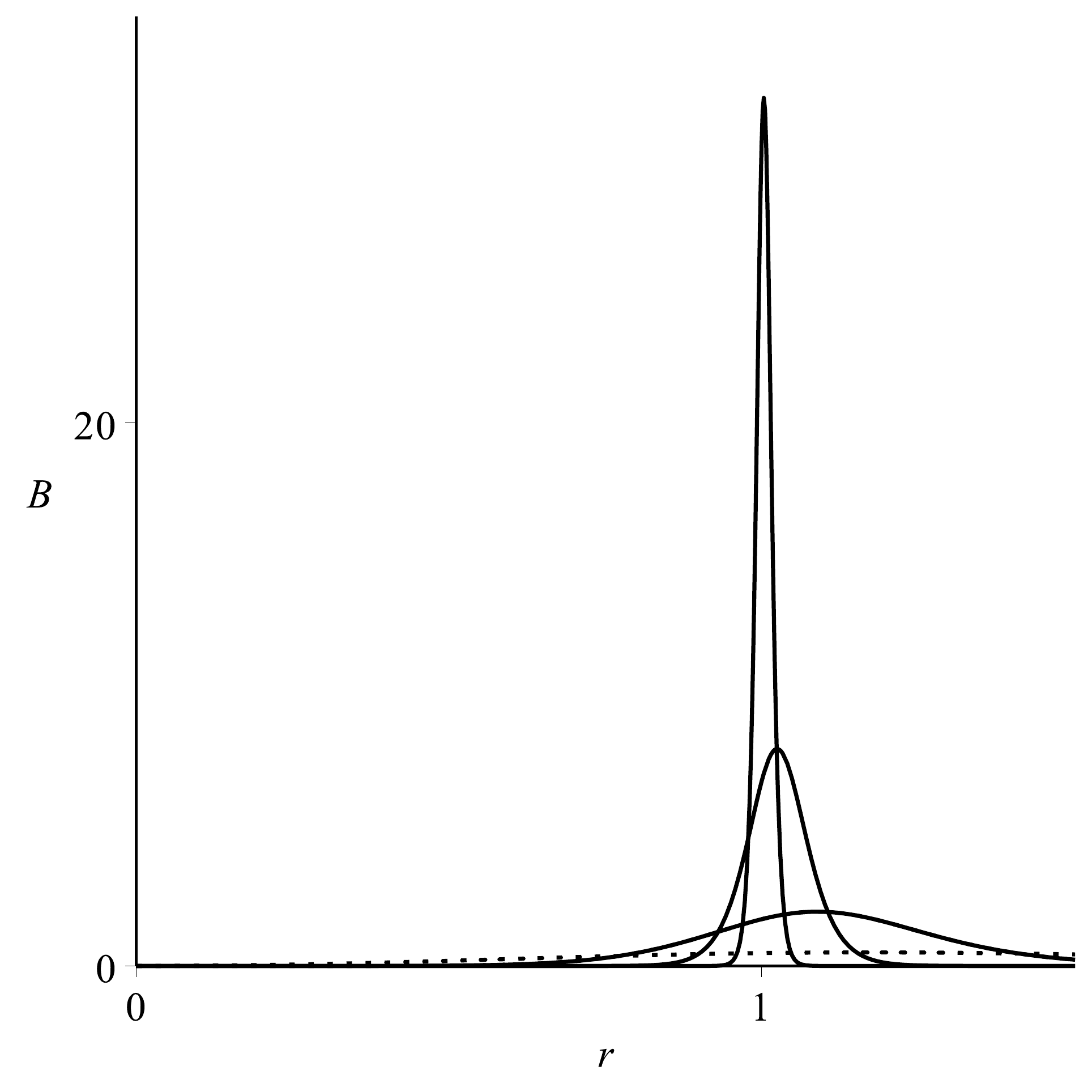}
\includegraphics[width=4.26cm]{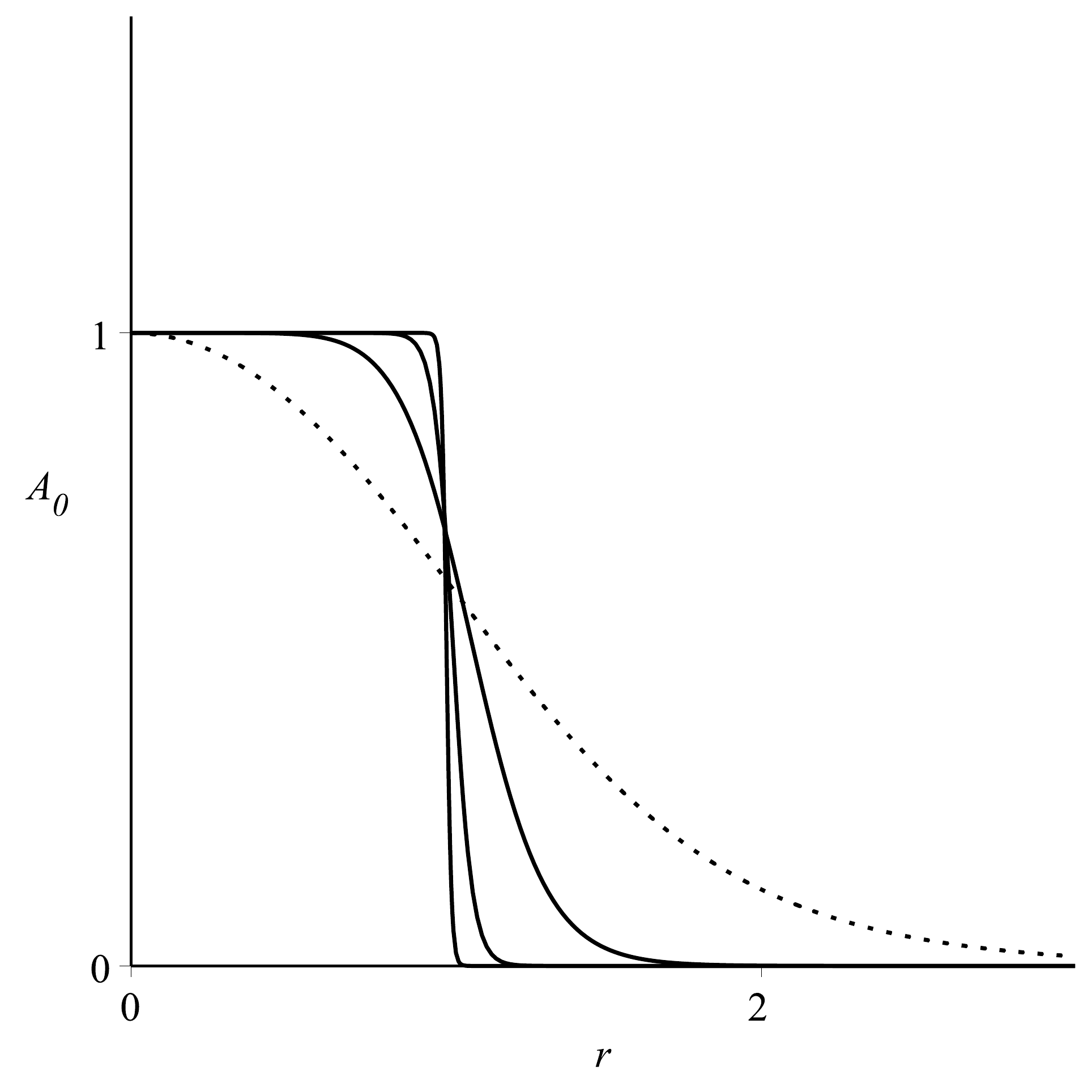}
\includegraphics[width=4.26cm]{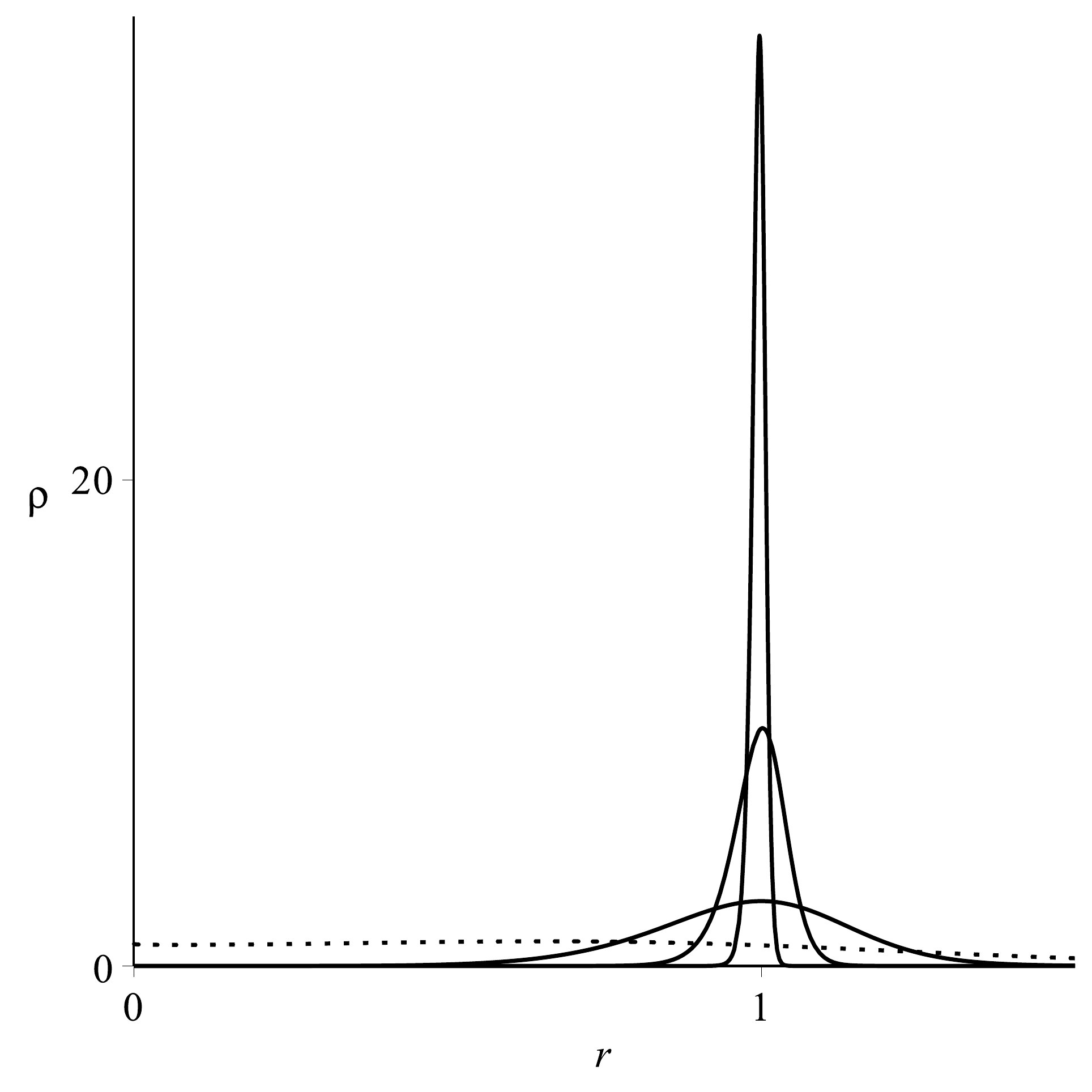}
\caption{The electric (top left) and magnetic (top right) fields \eqref{ebcs}, the temporal component of the gauge field \eqref{A0} (bottom left) and the energy density \eqref{rhomat} (bottom right), plotted for $l=1, 4, 16$ and $64$, with the case $l=1$ as the dotted line.}
\label{figebm}
\end{figure}
%%%%%%%%%%%%%%%%%

Also, we have to look at the electric and magnetic fields of Eq.~\eqref{ebcs} to study its behavior. In Fig.~\ref{figebm}, we have sketched the electric and magnetic fields, as well as the temporal gauge field $A_0$ of Eq.~\eqref{A0}, and the corresponding energy density. The uniform behavior of the functions $a_c(r)$ and $g_c(r)$ of Eq.~\eqref{solcmat} inside the compact sector makes the electric and magnetic fields tend to become a Dirac's delta as $l$ increases. For $l$ very large, these fields only exists as a very marrow ring around the unit value of $r$. Also, we study the energy density, given by Eq.~\eqref{rhoans}, which becomes
\be\label{rhomat}
\rho = \frac{{a^\prime}^2}{4r^2 lg^{2l}} + lg^{2l-2}{g^\prime}^2 +\frac{la^2g^{2l}}{r^2} + g^{2l}\left(1-g^{2l}\right)^2.
\ee
It is also displayed in Fig.~\ref{figebm} for several values of $l$. 

We see that as $l$ increases, the electric field, the magnetic field and the energy density become taller and thinner. They tend to become compact, living in a narrow area around $r=r_c$. In particular, the energy density behaves in a way such that the total energy obtained by numerical integration is always approximately equal to $2\pi$, regardless the value of $l$. We then conclude that the vortex tends to become compact as $l$ increases to larger and larger values, existing only inside a narrow ring around $r=r_c$.

\begin{figure}[t!]
\centering
\includegraphics[width=4.2cm]{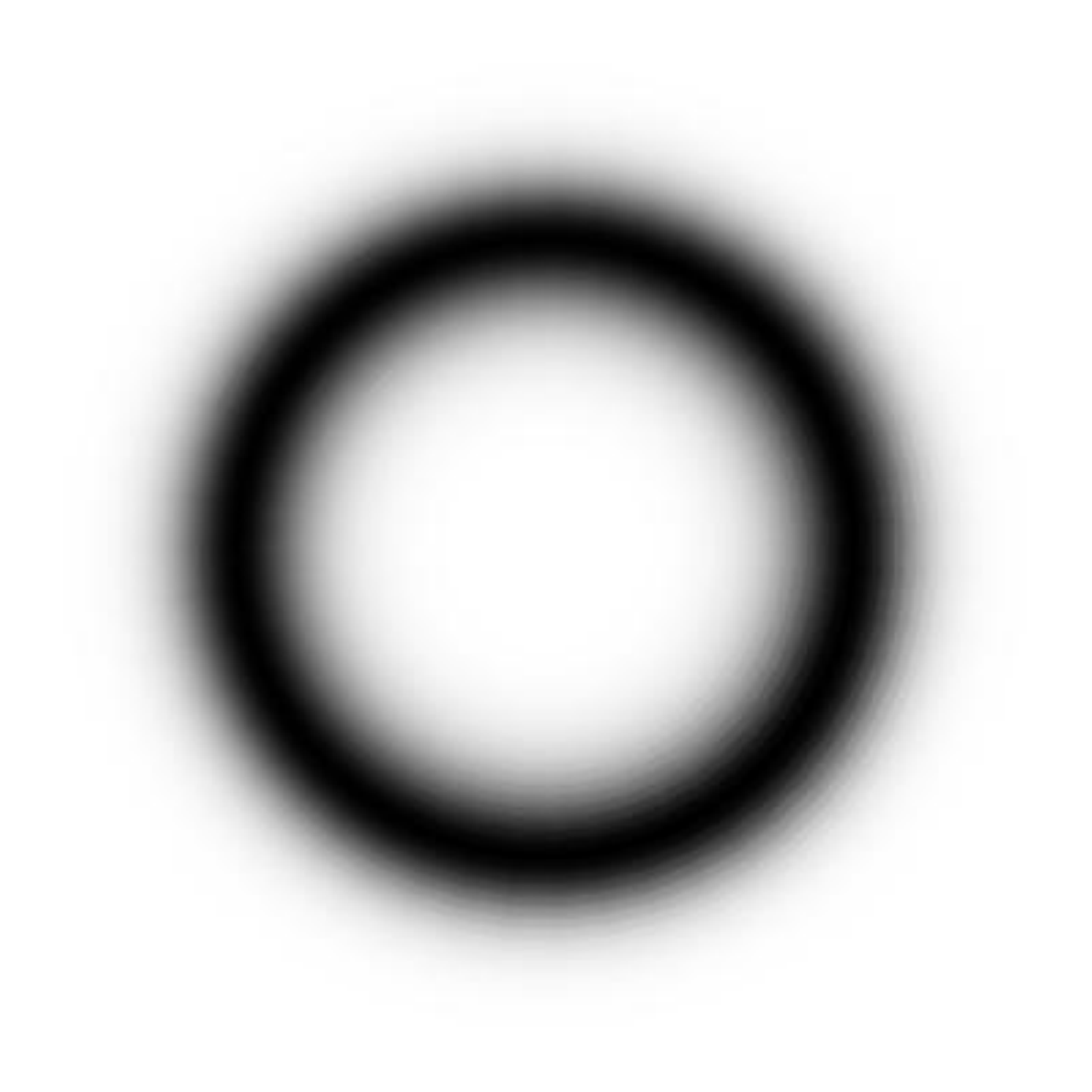}
\includegraphics[width=4.2cm]{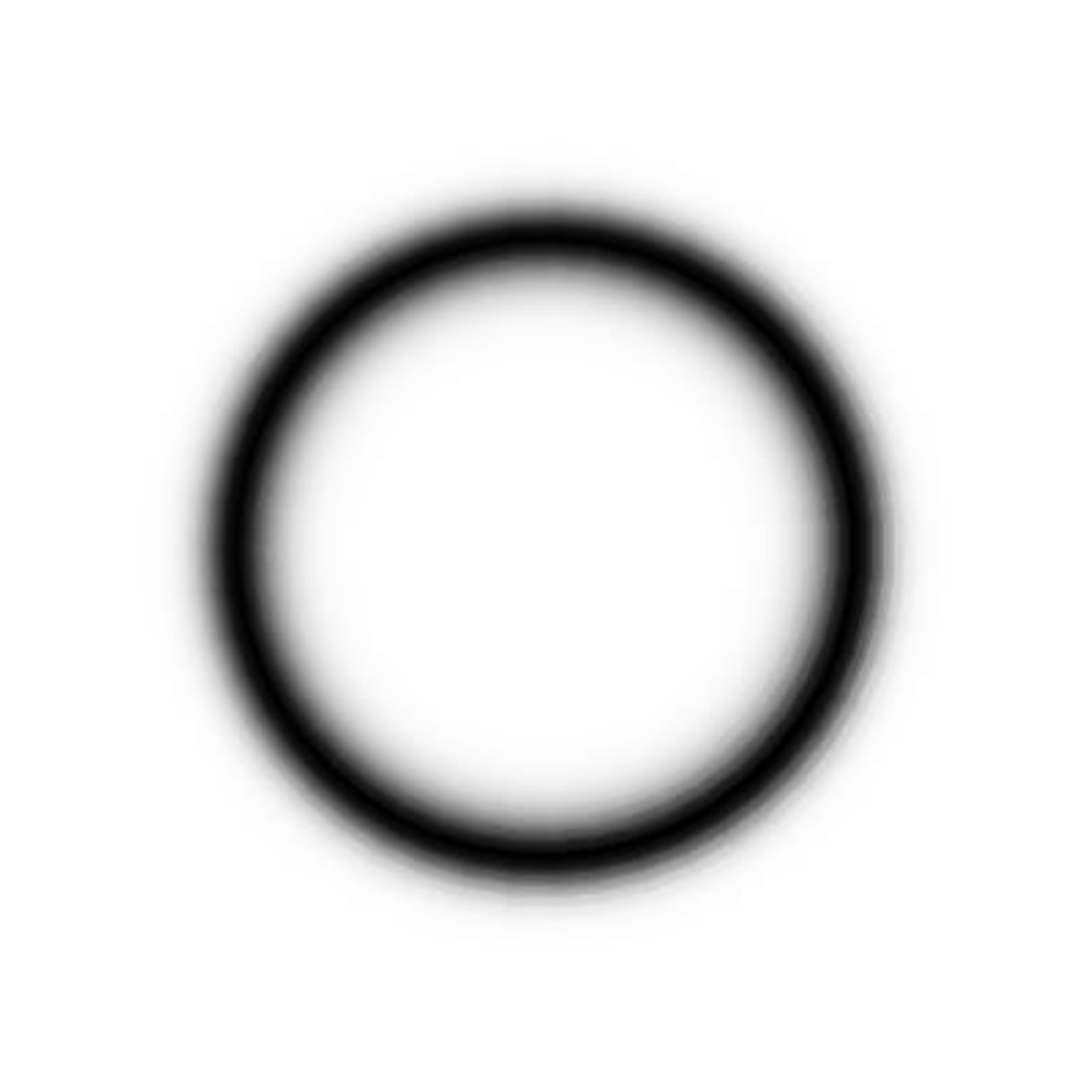}
\includegraphics[width=4.2cm]{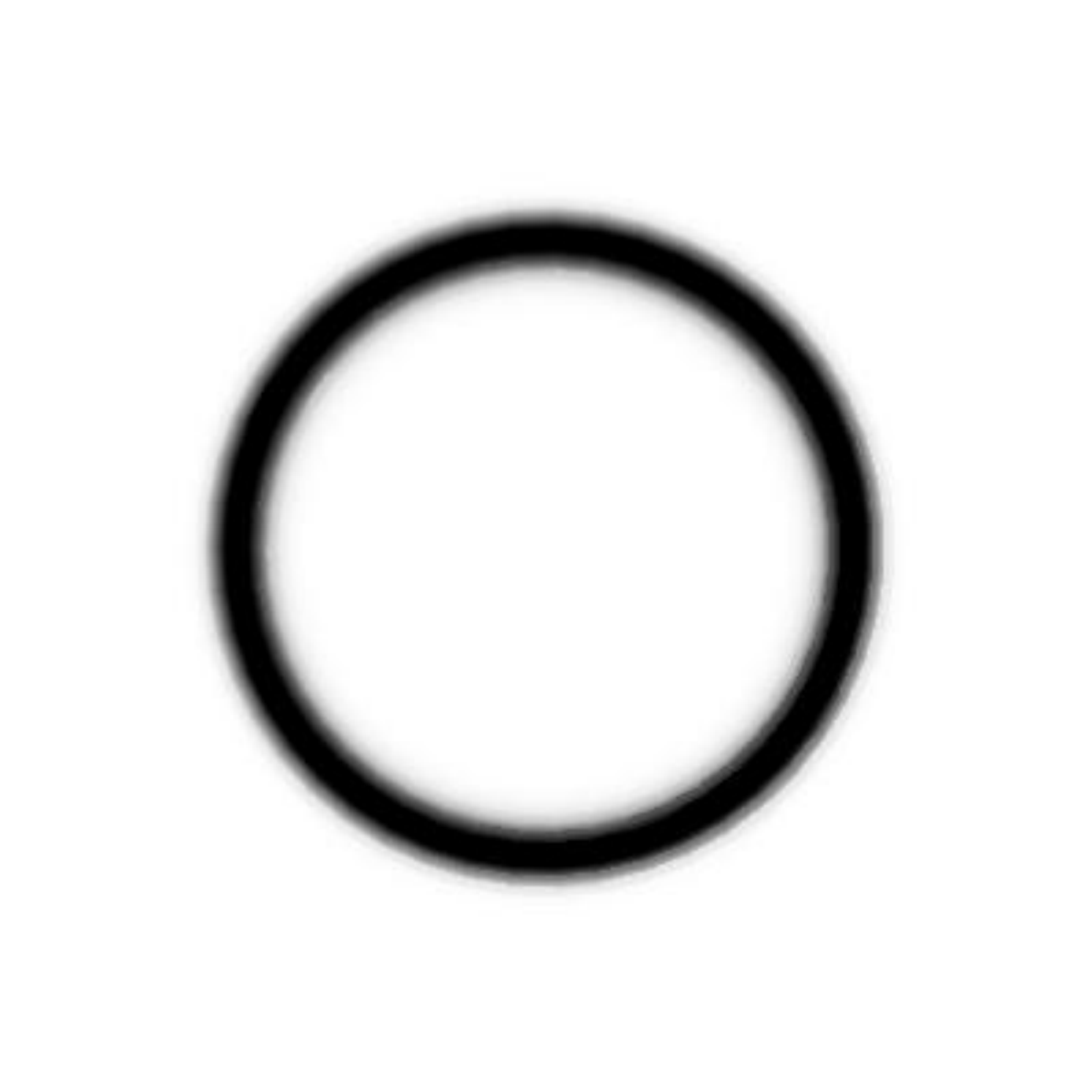}
\includegraphics[width=4.2cm]{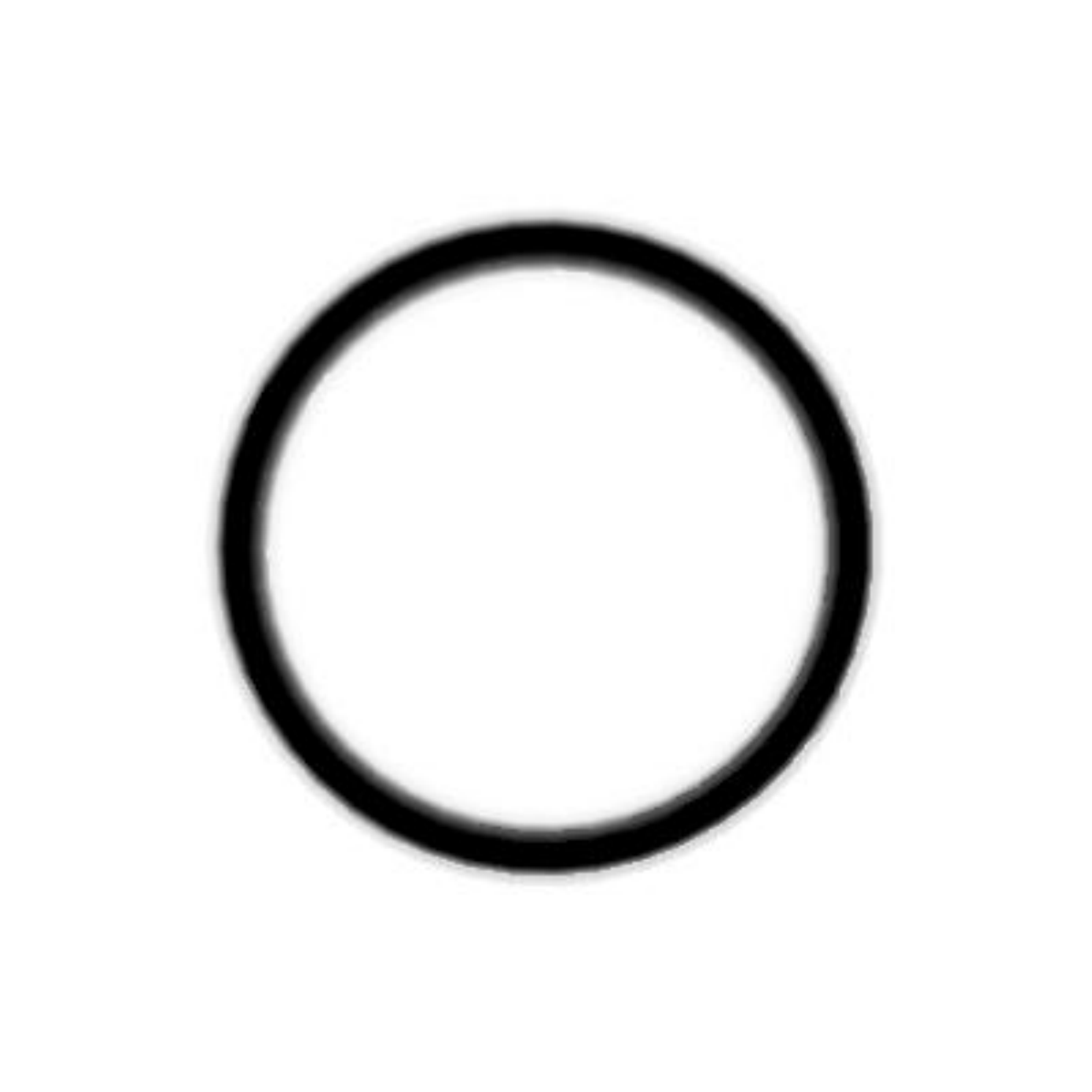}

\caption{The energy density \eqref{rhomat} depicted in the polar coordinates $(r,\theta)$ for $l=4$ (top left), $l=8$ (top right), $l=16$ (bottom left) and $l=32$ (bottom right), illustrating that the vortex shrinks to become compact, living inside a narrower and narrower ring around $r=r_c$ as $l$ increases to larger and larger values.}
\label{fig7}
\end{figure}

As far as we can see, this behavior was never seem before, so we illustrate it again in Fig.~\ref{fig7} where one depicts the energy density using polar coordinates, for some values of $l$. 

%%%%%%%%%%%%%%%%%%%%%%%
%%%%%%%%%%%%%%%%%%%%%%

%%%%%%%%%%%%%%%%%%%%
\section{Conclusions}
\label{conclusions}

In this work we studied two new models described by generalized Chern-Simons systems of the class \eqref{lcomp}. The two models are controlled by a single parameter that appears in the potential and in the function $K(|\vphi|)$ that modifies the scalar field dynamics. In the first model, we showed that although the solutions tend to go to their boundary conditions faster when the parameter gets larger, they do not shrink enough  to make the vortices compact. This happens because the potential tends to acquire the very same form of the one that appears in the standard case presented before in Refs.~\cite{cs1,cs2}. 

The investigation continued with the study of another model, the second model which is described by a potential whose concavity around its unit minimum tends to become narrower as the parameter increases to larger and larger values. This behavior is illustrated in Fig.~\ref{figVm} and is very nice, since it makes the solutions shrink significantly, inside a narrow ring around $r=r_c$, as illustrated in Figs.~\ref{figebm} and \ref{fig7}. This model is very different from the previous one and we could see that it starts with the standard model described in Refs.~\cite{cs1,cs2} in the case $l=1$. The effect of the compactification engendered by the second model is original and very interesting, and since it appears in a system that supports first-order differential equations, it may perhaps be extended to become supersymmetric and so of more general interest.

Other issues concern modification of the current work to consider the case of non-relativistic dynamics \cite{jp} and also, extensions of the Abelian $U(1)$ symmetry to the non-Abelian case. 

\acknowledgements{We would like to thank the Brazilian agency CNPq for partial financial support. DB thanks support from fundings 455931/2014-3 and 306614/2014-6, LL thanks support from fundings 307111/2013-0 and 447643/2014-2, MAM thanks support from funding 140735/2015-1, and RM thanks support from fundings 455619/2014-0 and 306826/2015-1.}

%%%%%%%%%%%%%%%%%%%%%%%%%%%%


\begin{thebibliography}{99}
\bibitem{novortex} H.B. Nielsen and P. Olesen, Nucl. Phys. \textbf{B61}, 45 (1973).
\bibitem{vega} H.J. de Vega and F.A. Schaposnik, Phys. Rev. \textbf{D14}, 1100 (1976).
\bb{cs}S.-S. Chern and J. Simons, Ann. Mathematics, {\bf99}, 48 (1974).
\bb{jac}S. Deser, R. Jackiw, and S. Templeton, Phys. Rev. Lett. {\bf48}, 975 (1982); Ann. Phys. {\bf140}, 372 (1982).
\bb{csp1} C.R. Hagen, Ann. Phys. {\bf157}, 342 (1984); Phys. Rev. D {\bf31}, 2135 (1985).
\bb{cs1} J. Hong, Y. Kim, P.Y. Pac, Phys. Rev. Lett. {\bf64}, 2230 (1990).
\bb{cs2} R. Jackiw, E.J. Weinberg, Phys. Rev. Lett. {\bf64}, 2234 (1990).
\bb{cs3} R. Jackiw, K. Lee, E.J. Weinberg, Phys. Rev. D {\bf42}, 3488 (1990).
\bb{dunne}G. Dunne, {\it Self-dual Chern-Simons theories.} (Springer-Verlag, 1995).
\bb{gen1}J. Lee and S. Nam, Phys. Lett. B {\bf261}, 437 (1991).
\bb{gen2}D. Bazeia, Phys. Rev. D {\bf46}, 1879 (1992).
\bb{kinf} C. Armendariz-Picon, T. Damour, V. Mukhanov, Phys. Lett. B {\bf458}, 209 (1999).
\bb{cosm1} C. Armendariz-Picon, V. Mukhanov, Paul J. Steinhard, Phys. Rev. Lett. {\bf85}, 4438 (2000). 
\bb{cosm2} C. Armendariz-Picon, V. Mukhanov, Paul J. Steinhardt, Phys. Rev. D {\bf63}, 103510 (2001).
\bb{babichev1} E. Babichev, Phys. Rev. D {\bf74}, 085004 (2006).
\bb{babichev2} E. Babichev, Phys. Rev. D {\bf77}, 065021 (2008).
\bb{kd1} X. Jin, X. Li, D. Liu, Class. Quantum Grav. {\bf24}, 2773 (2007).
\bb{kd2} S. Sarangi, JHEP {\bf018}, 0807 (2008).
\bb{kd3} D. Bazeia, L. Losano, R. Menezes, J.C.R.E. Oliveira, Eur. Phys. J. C {\bf51}, 953 (2007).
\bb{kd4} D. Bazeia, L. Losano, R. Menezes, Phys. Lett. B {\bf668}, 246 (2008).
\bb{kd5} D. Bazeia, A.R. Gomes, L. Losano, and R. Menezes, Phys. Lett. B {\bf671}, 402 (2009).
\bb{kd6} R. Casana, M.M. Ferreira, Jr, and E. da Hora, Phys. Rev. D {\bf86}, 085034 (2012).
\bb{kd7} Handhika S. Ramadhan, Phys. Lett. B {\bf758}, 149 (2016).
\bb{cas}R. Casana, M. L. Dias, and E. da Hora, Phys. Lett. B {\bf768}, 254 (2017). 
\bb{rosenau} P. Rosenau and J.M. Hyman, Phys. Rev. Lett. {\bf70}, 564 (1993).
\bb{ck1} D. Bazeia, L. Losano, M.A. Marques, and R. Menezes, Phys. Lett. B {\bf 736}, 515 (2014).
\bb{ck2} D. Bazeia, L. Losano, M.A. Marques, and R. Menezes, EPL  {\bf 107}, 61001 (2014).
\bb{ck3} D. Bazeia, M.A. Marques, and R. Menezes, EPL  {\bf 111}, 61002 (2015).
\bb{ck4} D. Bazeia, L. Losano, M.A. Marques, R. Menezes, and R. da Rocha, Phys. Lett. B {\bf 758}, 146 (2016).
\bb{ckg} D. Bazeia, L. Losano, and R. Menezes, Phys. Lett. B {\bf 731}, 293 (2014).
\bb{cvmh} D. Bazeia, L. Losano, M.A. Marques, R. Menezes, and I. Zafalan, Eur. Phys. J. C {\bf 77}, 63 (2017).
\bb{bazeiacs}D. Bazeia, E. da Hora, C. dos Santos, and R. Menezes, Phys. Rev. D {\bf 81}, 125014 (2010).
\bb{jp}R. Jackiw and S.-Y. Pi, Phys. Rev. Lett. {\bf64}, 2969 (1990); Phys Rev D {\bf42}, 3500 (1990).
\end{thebibliography}
\end{document}